\documentclass[aps,groupedaddress,superscriptaddress,prd,twocolumn,nofootinbib]{revtex4}

\usepackage{amsmath,amssymb,bm}
\usepackage{graphicx}
\usepackage{epstopdf}
\usepackage{amsfonts}
\usepackage{amssymb}
\usepackage{amsbsy}
\usepackage{amsmath}
\usepackage{latexsym}
\usepackage{sansmath}
\usepackage{subfigure}
\usepackage{lipsum}
\usepackage{float}
\usepackage{bm}
\usepackage{color}
\usepackage{comment}
\usepackage{csquotes}								
\usepackage{physics}
\usepackage{eufrak}
\usepackage{accents}
\usepackage[colorlinks=true,breaklinks]{hyperref}

\usepackage[plain]{algorithm}
\usepackage{algpseudocode}

\def\be{\begin{equation}}
\def\ee{\end{equation}}
\def\bea {\begin{eqnarray}}
\def\eea {\end{eqnarray}}
\def\nn {\nonumber}
\def \p {\partial}

\begin{document}
 
\title{Quantum backreaction on a classical universe}
\author{Viqar Husain} \email{vhusain@unb.ca} 
\affiliation{Department of Mathematics and Statistics, University of New Brunswick, Fredericton, NB, Canada E3B 5A3}
 \author{Suprit Singh} \email{suprit@iitd.ac.in} 
\affiliation{Department of Mathematics and Statistics, University of New Brunswick, Fredericton, NB, Canada E3B 5A3}
\affiliation{Department of Physics, Indian Institute of Technology Delhi, Hauz Khas, New Delhi, India 110016} 

\begin{abstract}
\vskip 0.2cm
We study a  first-order formulation for the coupled evolution of a quantum scalar field and a classical Friedmann universe.  The model is defined by a state dependent hamiltonian constraint and the time dependent Schr\"odinger equation for the scalar field.  We solve the resulting non-linear equations numerically for initial data consisting of a Gaussian scalar field state and  gravity phase space variables. This gives a self-consistent semiclassical evolution that includes non-perturbative ``backreaction" due to particle production. We compare the results with the evolution of a quantum scalar field on a fixed background, and find that the backreaction modifies both particle production and cosmological  expansion, and that these effects remain bounded.    
  
\end{abstract}

\maketitle

\section{Introduction}
   
 Semiclassical gravity is  the study of quantum matter propagating on a classical spacetime. One part of this broadly defined area is the study of quantum fields in curved spacetime (QFTCS), a subject that has been extensively studied (see textbook refs.~\cite{birrell_davies_1982, fulling_1989,mukhanov_winitzki_2007,Parker:2009uva}). The other and less well-established field is the extension of QFTCS to include back reaction of  quantum stress-energy on classical spacetime. This paper concerns the latter. 
  
 There are at present two broad approaches for addressing the problem of semiclassical gravity with back reaction: the first and earliest is the  postulation of a semiclassical Einstein equation without first having at hand a quantum theory of gravity. The proposal 
 \be
 G_{ab} = 8\pi G \langle \Psi | \hat{T}_{ab}|\Psi \rangle
 \label{G=T}
 \ee 
has been the focus of much research, see e.g. \cite{Moller:1962aa,Utiyama:1962aa,Parker:1969aa,DEWITT1975295, Page:1981aj}. For  a recent review see e.g. \cite{Ford:2005qz}. 
 
 The second is to start from a proposal for quantum gravity and make a suitable ansatz for geometry and matter physical states to arrive at a consistent approximation.  This typically starts with the Dirac quantization conditions which give the Wheeler-deWitt equation and the diffeomorphism conditions
 \bea
 \left(\hat{H}^G + \hat{H}^M\right) |\Psi\rangle =0,\label{wdw} \\
 \left( \hat{C}^G_a  + \hat{C}^M_a \right) |\Psi\rangle =0.
 \label{diffeo}
 \eea
To arrive at a semi-classical approximation from these equations, the joint gravity-matter state is carefully chosen such that the gravity state is semiclassical and peaked on a classical configuration, and the matter quantum state is parametrized by the  classical peaking configuration. The combined state is a product of such matter and gravity states. The Wheeler-DeWitt equation than simplifies to a time-dependent (functional) Schrodinger equation.  
  
 There has been a considerable work in trying to extract (\ref{G=T}) from the Dirac quantization (\ref{wdw})-(\ref{diffeo}) of gravity with matter \cite{Halliwell:1987eu,Boucher:1988ua,Padmanabhan:1989aa,Singh:1989aa,Kiefer:1990pt,Kiefer:1993fg,Kiefer:2014sfr,Kiefer:2018xfw}. Beyond mini-superspace models the results are largely formal in nature. This is because the problem of preserving the quantum algebra of constraints, a crucial ingredient for maintaining spacetime reparametrization invariance in the quantum theory, remains unsolved.  
  
A related observation and potential problem of the proposal (\ref{G=T}) is in its very definition. In the  general setting, this is an equation for a semiclassical metric $g$ given a fixed state $|\Psi\rangle$ in the Heisenberg representation. However, viewed as a non-perturbative equation, the state and the operator $\hat{T}_{ab}(\hat{\phi}, g)$ must be defined with respect to the  as-yet-undetermined metric $g$. But the field modes used to define $\hat{T}_{ab}$ are normally solutions of the wave equation, which in turn requires a metric for their definition. This situation is  of course unlike the case of QFT on a fixed background because there a background metric is available and provides the necessary  mode equation; its solutions can be used to define (up to coordinate choices) the Heisenberg operator $\hat{T}_{ab}(\hat{\phi},g)$. There is, thus, the question of exactly how the r.h.s. of (\ref{G=T}) is to be constructed if the metric is not known explicitly. This is an issue even for the simplest cosmological metrics where the only free function is the scale factor $a(t)$: there is no analytical solution of the wave equation for a general form of $a(t)$, so the r.h.s. of  (\ref{G=T}) cannot be constructed. Several additional issues with (\ref{G=T}) are outlined in refs. \cite{Isham:1995wr, Ford:1997hb}.  
  
For spherically symmetric and cosmological solutions of Einstein equations with no free functions in the metric there is a large literature on calculations of $\langle \hat{T}_{ab}\rangle$ for various choices of vacuum states; these fall in the domain QFTCS, and can be used to define (\ref{G=T}).  An example of this is the replacement of the Schwarzschild mass $M$ by $M(t)$ to model a black hole that is shrinking due to Hawking radiation. These simple cases can be compared with the mini-superspace models of quantum gravity where all metric functions also depend only on a time parameter.    
 
These considerations raise the broader question of how to define a coupled classical-quantum system. There have been some efforts to incorporate  semi-classical back reaction on a classical system, e.g. \cite{Kuo:1993if,Anderson:1995tn,Brout:1995aa,Yang:2012mh, Levasseur:2014ska,Struyve:2015cwa,Tilloy:2015zya,Vachaspati:2017jtw,Bojowald:2020emy,Sudarsky2021}. These attempts essentially fall into two categories, one of which tries to modify the dynamics on either side (classical or quantum) in exotic ways (from discarding unitarity to introducing new structures such as stochastic fluctuations in the classical sector). The second class either seeks to recover ($\ref{G=T}$), but typically fails to incorporate the full canonical structure self-consistently. 

With this broad motivation, we address here the general question of how to couple a quantum theory to a classical theory in the first order formalism such that initial data consisting of a classical configuration and a quantum state evolve self-consistently.  In Sec. II we outline the approach by considering a scalar field coupled to a flat Friedmann-Lemaitre-Robertson-Walker (FLRW) cosmology where the scalar field is quantized. We next apply the general idea to backreaction due to particle production is cosmology: in Sec. III, as a prelude we describe a canonical approach to particle creation in a fixed FLRW cosmology, where we show, unlike in the standard covariant asymptotic  computation, that the particle number in each mode varies  in time as it reaches saturation at late time. In Sec. IV we self-consistently calculate the scale factor and particle number with backreaction. We  show that this results is a modified expansion rate that is more pronounced at early times. In Section V we close with a summary and prospects for the applications of our method in other gravitational settings where back reaction is expected to play an important role. 
     
\section{Self-consistent quantum-classical cosmology}
   
 We describe here our method for the joint evolution of a quantum-classical system in cosmology. The classical system is a flat FRLW cosmology and the quantum system is  a scalar field. The equations we present may be viewed as ones that describe backreaction non-perturbatively. However this characterization is not appropriate because the term "backreaction" presumes an apriori fixed background spacetime with a test quantum field, which then subsequently acts to deform the spacetime as a higher order effect.  As we will see, the equations we propose are such that a classical spacetime and matter quantum state evolve self-consistently from initial data in a manner that does not permit identification of  an apriori classical background spacetime.  
   
 Although we do not consider inhomogeneities, it will be evident that our  method is  different in both spirit and technique from the ``backreaction" method deployed for cosmological perturbation theory where the effective dynamics of the perturbations is, in the final analysis,  equivalent to a scalar field on the a fixed FRLW background. In our approach there is no fixed background  that forms the basis for a perturbation expansion. 

The canonical variables for this model are the scale factor and scalar field and their conjugate momenta $(a,p_a)$ and $(\phi,p_\phi)$. The metric is 
\be
ds^2 = -N^2dt^2 + a^2(t)(dx^2 + dy^2 + dz^2)
\ee
 and the classical dynamics is described by the Hamiltonian constraint (for $\Lambda =0$)  
\be
{\cal H} = -\frac{p_a^2}{24a}+\frac{p_\phi^2}{2 a^3}+ a^3 V(\phi) \equiv -\frac{p_a^2}{24a} + h_\phi,  
\ee
and the Hamilton equation that follow from it, 
\bea
\dot{a} &=& \{a , N{\cal H}\}, \quad \dot{p}_a = \{p_a, N{\cal H}\},\\
\dot{\phi} &=& \{\phi , N{\cal H}\}, \quad \dot{p}_\phi = \{p_\phi, N{\cal H}\}.
\eea  

We define the hybrid quantum-classical theory by quantizing  the scalar field  and defining an effective state dependent Hamiltonian constraint 
\be
{\cal H}_\Psi \equiv -\frac{p_a^2}{24a}+ \frac{1}{2a^3} \langle \Psi | \hat{p}_\phi^2 |\Psi\rangle + a^3 \langle \Psi  |V(\hat{\phi})|\Psi\rangle =0,
\label{heff}
\ee
where $|\Psi\rangle$ is any state of the scalar field. The proposed evolution equations (choosing lapse $N=1$) are the coupled set
\bea
\dot{a} &=& \{a , {\cal H}_\Psi\} = -\frac{p_a}{12a}, \label{aeef}\\
 \dot{p}_a &=& \{p_a, {\cal H}_\Psi\} \nn\\
                 &=&  -\frac{p_a^2}{24a^2} +\frac{3}{2a^4} \langle \hat{p}_\phi^2 \rangle_\Psi 
                 -3a^2  \langle V(\hat{\phi}) \rangle_\Psi \label{paeff}\\ 
i  |\dot{\Psi}\rangle &=& \frac{1}{2a^3} \ \hat{p}_\phi^2 |\Psi\rangle + a^3 V(\hat{\phi}) |\Psi\rangle
\label{tdse0}
\eea
where we are working in the Schrodinger picture. Initial data for these equations is $\{a(t_0),p_a(t_0),|\Psi\rangle(t_0) \}$ chosen such that the constraint (\ref{heff} ) holds. In practice this means solving the quadratic for $p_a(t_0)$ given $a(t_0)$ and $|\Psi\rangle(t_0)$. 
The first of these equations is the same form as the corresponding classical one with the difference that $p_a$ on the r.h.s. is state dependent through the second equation. 

For self-consistency we must verify that the constraint ${\cal H}_\Psi =0$ is conserved. This is checked by using the evolutions (\ref{aeef})-(\ref{tdse0}):
\bea
\frac{d}{dt} {\cal H}_\Psi  &=& \frac{\p {\cal H}_\Psi }{\p a} \ \dot{a} + \frac{\p {\cal H}_\psi }{\p p_a }\  \dot{p}_a  
                           + \langle \dot{\Psi}| \hat{h}_\Phi | \psi\rangle + \langle \Psi | \hat{h}_\phi | \dot{\Psi}\rangle \nn\\
                           &=&0.                          
\eea 
 This completes the prescription for a first order formulation of the hybrid classical-quantum cosmology. The initial scalar field state is arbitrary; it can be a linear combination Gaussian or squeezed states, or indeed any normalizable state. The resulting cosmological evolution is much richer than in the  purely classical theory since there is an arbitrary normalizable function's worth of initial data instead of the classical pair $(\phi(t_0), p_\phi(t_0)) \in \mathbb{R}^2$. Solutions to this system of equations may be generated numerically \cite{Husain:2018fzg}.
 
 In the next section we apply this approach to calculate the self-consistent evolution of a Gaussian state of the scalar field for all modes $k$, and determine the corresponding evolution of the classical cosmology. 
    
   \section{Canonical approach to particle production}
   
  As a prelude to the main calculation in the next section using the ideas we outlined above, we revisit here the cosmological particle production  calculation. We redo this calculation using the time-dependent Schrodinger equation (TDSE) for each mode of a scalar field on a background prescribed by a scalar factor $a(t)$.  
   
   The standard calculation proceeds as follows: one solves the scalar wave equation for the early and late time modes $f_E(k)$ and $f_L(k)$. These two sets of modes define distinct  bases for the Fock space, together with their corresponding number operators $N_E$ and $N_L$. Cosmological particle production is result that the expectation value of the late time mode number operator $N_L$ in the vacuum $|0_E\rangle$ of the early time modes is non-zero, i.e. $\langle 0_E| N_L | 0_E\rangle \ne 0$.   

 The TDSE method starts from the Hamiltonian of a scalar field of mass $m$ on a given FRLW background with scale factor $a(t)$. In spatial Fourier space with modes $k= |\vec{k}|$, the Hamiltonian is 
\bea
  H &=& \frac{1}{2}  \int \frac{d^3k}{(2\pi^3)}\left[  \frac{1}{ a^3}\  p_k^2 +  a^3 \left(\frac{k^2}{a^2}  + m^2\right)\phi_k^2     \right], \nn\\
     &\equiv& \int \frac{d^3k}{(2\pi^3)}\  h_k,
     \label{hkk}
\eea  
where $p_k$ and $\phi_k$ are defined from the Fourier transforms of the scalar field and its conjugate momentum.

The quantization of each mode may be carried out in the standard manner leading to the TDSE 
\be
i \frac{\p}{\p t}\Psi_k(\phi_k, a(t),t) = \hat{h}_k \Psi_k (\phi_k,a(t),t).
  \label{k-tdse}
\ee

The method for calculating cosmological particle production from this equation is the following. First we solve (\ref{k-tdse}) using the evolving Gaussian ansatz
\be
  \Psi_k (\phi_k,t) = \beta_k(t) \exp \left[-\alpha_k(t) \phi_k^2 \right],\label{gauss}
  \ee 
  where $\alpha_k$ and $\beta_k$ are complex values functions of $t$. At the chosen initial time, this state coincides with the instantaneous eigenstate of $\hat{h}_k$.  The TDSE with this ansatz leads to the  equations 
 \bea
  i\ \frac{d \alpha_k}{dt} &=& \frac{1}{2a}\left( 4\alpha^2_k -k^2\right),
 \label{alphadot}\\
 i \frac{d}{dt}\left(\ln \beta_k\right)  &=& \frac{\alpha_k}{a}.   
 \eea
 Normalization gives 
 \be
  |\beta_k|^2 = \sqrt{\frac{2\Re( \alpha_k)}{\pi}}.
 \ee

To compute the particle number in mode $k$ at time $T>t_0$, we note that the eigenvalue problem for $\hat{h}_k$ (\ref{hkk}) determines an instantaneous basis $\psi_n^k(\phi_k,T)$ for the Hilbert space at $T$. And since $\hat{h}_k$ is an oscillator with mass $a^3$ and frequency $m^2 + k^2/a^2$, $\psi_n^k(\phi_k,T)$ are just the oscillator eigenfunctions. The overlap $| \left (\Psi_k(T), \psi_n^k(T) \right)|^2$ of the evolved wave function  $ \Psi_k (\phi_k,T)$ with such a  basis element gives the probability that the evolved state of mode $k$  has instantaneous excitation level $n$ at time $T$. Hence the particle number in mode $k$ at time $T$ is 
 \be
  \langle n_k(T)\rangle = \sum_{n=0}^\infty \ n\ | \left (\Psi_k(T), \psi_n^k(T) \right)|^2.
  \label{nk}
  \ee
  \begin{figure} [ht]
\centering
\includegraphics[width=8cm]{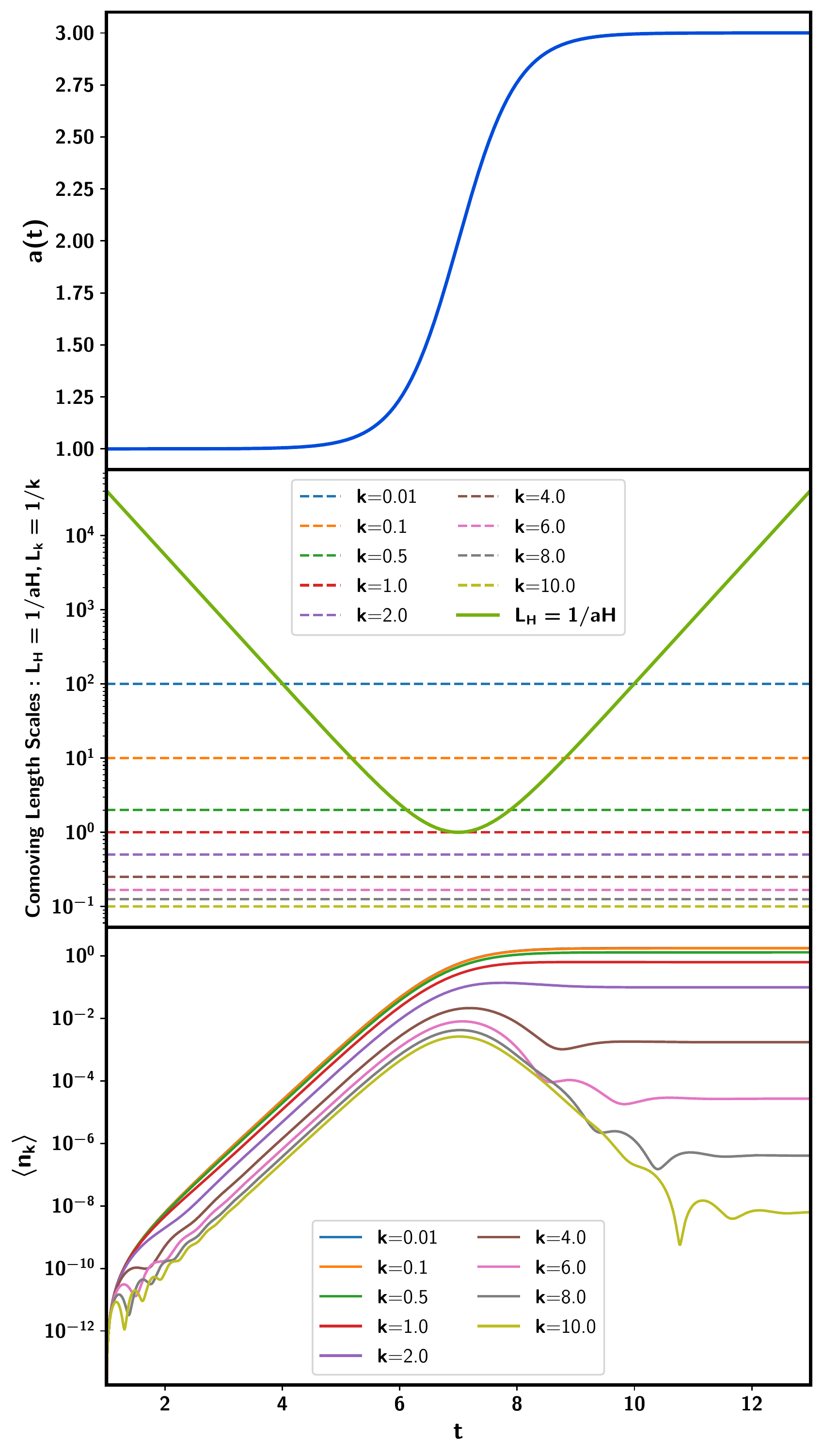}
\caption{Particle production by mode in the Schr\"odinger picture: the top frame is the scale factor used, the second illustrates the mode scales in relation to the Hubble scale, and the bottom one is the computed particle production for selected modes using eqn. (\ref{nk}). }
\end{figure}

This overlap integral and sum can be computed exactly with the result \cite{Mahajan:2007qc,Mahajan:2007qg}
  \be
\langle  n_k\rangle = \frac{|z_k|^2}{1-|z_k|^2}, 
\label{nkf}
   \ee 
   where $z_k(t)$ is defined from the solution $\alpha_k(t)$ of eqn. (\ref{alphadot}) by
   \be
    \alpha_k = \frac{ka^2}{2} \left( \frac{1-z_k}{1+z_k} \right).
   \ee 
 An advantage of this  method  is that the particle number $n_k(t)$ may be computed at any time during cosmological evolution. This is unlike the standard method based on field expansion in the mode solutions of the wave equation where only the asymptotic late time particle number is available.  
  
 We performed this calculation for fixed scale factor of the form 
 \be
 a(t) = A + B \tanh\left(\frac{t-\bar{t}}{t_0}\right);
 \label{tanh}
 \ee 
 The results for $A=2, B=1, \bar{t}=7, t_0=1$ appear in Fig 1.  The first frame is the scale  factor, the second  shows the Hubble and mode length scales, and the last frame shows $n_k(t)$ for selected values of $k$. A number of features are visible in the last frame: particle production number rises initially with small oscillations for larger $k$; as the horizon scale falls toward its minimum $(t=7)$, the super horizon mode numbers reach their late time steady state values, whereas the sub horizon mode numbers dip before reaching their much lower steady states.  
 
 In summary, the purpose of this section was to demonstrate the Schr\"odinger picture technique for computing particle production. We  now combine this method with the quantum-classical idea discussed in Sec. II   to compute self-consistently the evolution of  the scale factor with the quantized scalar field. Although non-perturbative, this calculation implicitly contains ``back reaction" due to particle creation.

 \section{``Back reaction" in cosmology}
 
 In the last section we considered cosmological particle production with a specified scale factor. In light of the discussion there we now present and solve a set of equations that determine self-consistent evolution of the scale factor by taking into account particle production. This calculation is in the spirit of the self-consistent evolving wave function model described in Sec.II. The key difference is that the expectation value of the energy density in the effective Hamiltonian constraint ${\cal H}_\Psi$ is integrated over all modes with a Planck cut-off, and this sum is as the evolving source in the constraint. 
 
 Considering again a scalar field propagating on the flat FRLW cosmology. The semiclassical equations we propose are  
 \bea
  i \frac{\p}{\p t}{\Psi_k}(\phi_k, a(t),t) &=& \hat{h}_k \Psi_k (\phi_k,a(t),t),\label{modetdse}\\ 
  {\cal H}_{\Psi} &\equiv& -\frac{p_a^2}{24 a} + a^3 \langle\rho\rangle_\Psi =0
  \label{H-semi},\\
  \dot{a} &=& \{a, {\cal H}_{\Psi} \} = - \frac{p_a}{12 a},
  \label{adot}
  \eea
  where
 \be
\langle \rho \rangle_\Psi =  \frac{1}{a^3}\int \frac{d^3k}{(2\pi)^3}  \langle \hat{h}_k \rangle_\Psi.
\label{<rho>}
\ee 
The first of these is the TDSE for each mode, the same as what we used in Sec. III; the second is the effective Hamiltonian  constraint, where the matter density is the expectation value of the scalar field energy density integrated over all modes
 (\ref{<rho>}); and the third is the Hamilton equation for the scale factor. Initial data for these equations is the set 
 \be
 \{a(t_0), \Psi_k(t_0) \}.
 \ee
 At each time step, the state is used to compute $\langle\rho \rangle_\Psi$, and the semiclassical hamiltonian constraint (\ref{H-semi}) is solved at the initial time to determine $p_a(t_0)$. Evolution is then determined by solving the coupled first order system (\ref {k-tdse}) and (\ref{adot}), and the process is iterated.  We note that the evolution equation for $p_a$ is not used; as shown in Sec. II, the same argument that  gives conservation of ${\cal H}_\Psi$ applies to each mode, and hence to the integration over modes. Thus one can use either the $a$ and $p_a$ equations which preserve the effective constraint ${\cal H}_\Psi$,  or use the $a$ equation and the effective constraint to solve the solve the system.  
 
  For the time dependent Gaussian state (\ref{gauss})  eqns. (\ref{modetdse}-\ref{adot})  become
   \bea
   \label{br-tdse}
   i\dot{\alpha}_k &=& \frac{2\alpha_k^2}{a^3} -  \frac{a^3}{2}\left(  \frac{k^2}{a^2} +  m_\phi^2\right) \\
    \left(\frac{\dot{a}}{a}\right)^2 &=&  \langle \rho \rangle_\Psi,
   \label{br-H}
   \eea
  and a calculation gives 
   \be
    \langle \hat{h}_k \rangle = \frac{1}{2\Re (\alpha_k) } \left[ \frac{|\alpha_k|^2}{a^3} + \frac{a^3}{4}\left( \frac{k^2}{a^2} + m_\phi^2  \right)  \right],
   \ee
with $ \langle \rho \rangle_\Psi$ given by (\ref{<rho>}). The particle number  in mode $k$  is again given by (\ref{nk}).  

  \begin{figure*} [ht]
\includegraphics[width=\textwidth]{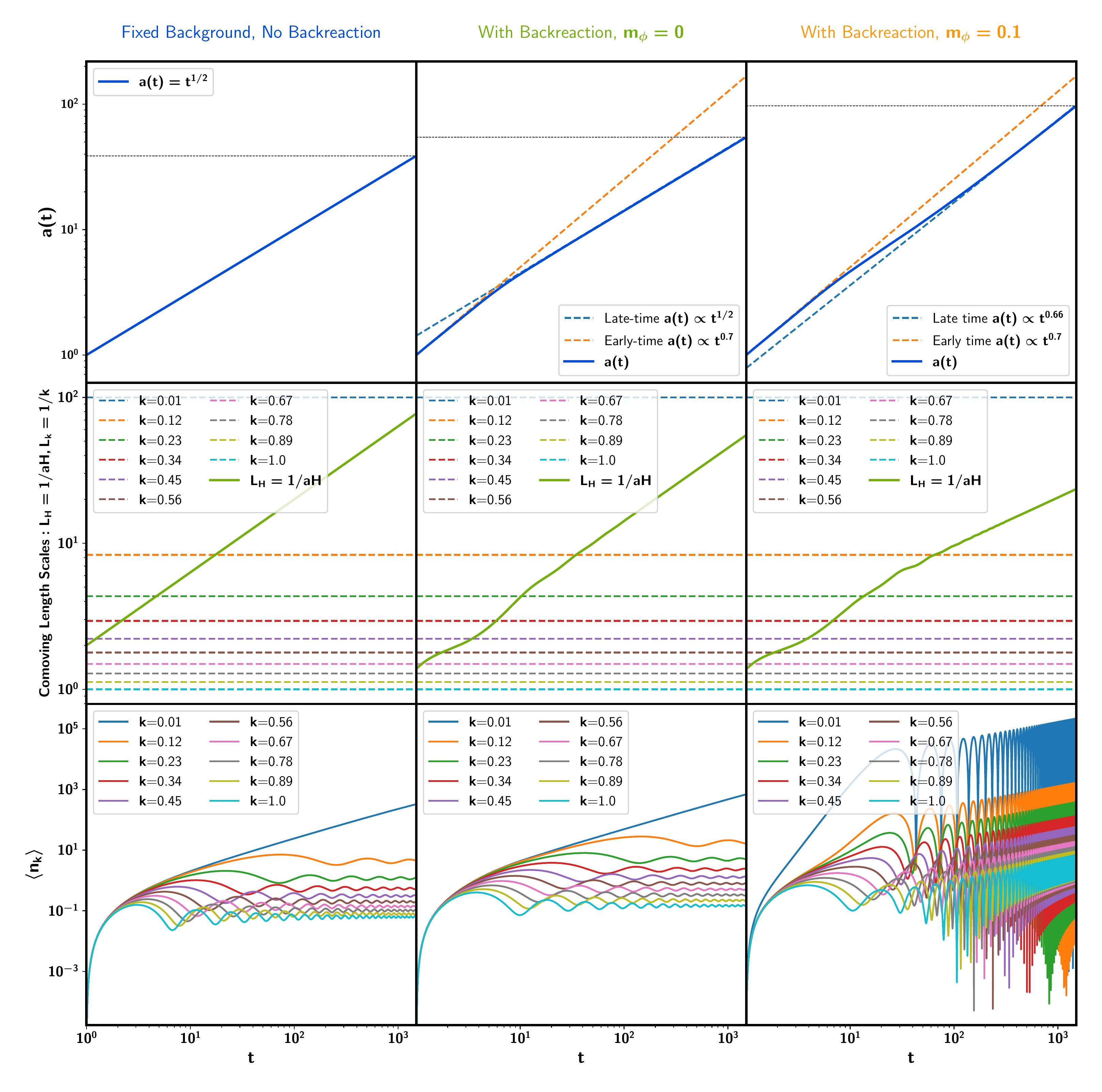}
\caption{Scale factor dynamics and particle production: the first column gives particle production without back reaction for $a(t)=t^{1/2}$. The second and third columns  are solutions of eqns. (\ref{br-tdse}-\ref{br-H}) that give the scale factor dynamics as modified by particle production; the dotted lines give the early and late time back reaction effects on the scale factor. The scales in the second row are in Planck units; $k=1$ is the Planck scale.}
\end{figure*}
 
We note that the zero mode $k=0$ is not included in the mode summed density due to the $k$-space volume factor in the computation of $\langle \rho \rangle_\Psi$. This may be added separately to the r.h.s of (\ref{br-H}), but we have not done so in the following calculations; this is the energy density  in the homogeneous model  in Sec. II.  For $k\ne 0$, the equations  represent the mode summed homogeneous contribution to the cosmological dynamics, and so exclude the explicit inhomogeneities that would require inclusion of the spatial diffeomorphism constraint.

    \subsection{Numerical solution}
  
  We computed the solutions of the coupled eqns. (\ref{br-tdse}-\ref{br-H}) for $a(t)$ and $\alpha_k(t)$, and used these to determine  $n_k(t)$ using (\ref{nk}). It is convenient for numerical integration to use dimensionless variables  defined by   rescaling with the Planck length $l_P$: $ t\rightarrow  t/l_P$, $\phi_k \rightarrow l_p \phi_k$, $\alpha_k \rightarrow \alpha_k/l_P^2$, $ k \rightarrow l_pk$, and $m_\phi\rightarrow l_p m_\phi$. For the density term in the Friedmann equation (\ref{br-H}) we used  the Planck scale ultraviolet cutoff $k=1$ (in Planck units) on the mode integration for the density,  namely
 \be
    \langle \rho\rangle_\Psi =  \frac{1}{a^3}\int_0^1 \frac{d^3 k}{(2\pi)^3}\ \langle\hat{h}_k \rangle_\Psi,  
  \ee
 
 Fig. 2. contains our results for initial data $a(0) =1$, and $\psi_k(0)$ given by the state  (\ref{gauss}) for a selection of modes $k$. The first column gives  particle production without back reaction for $a(t) = t^{1/2}$; this may be compared to
 Fig. 1 which gives the results of a similar calculation for the scale factor of eqn. (\ref{tanh}). The second and third columns give the self-consistent dynamics as modified by non-perturbative back reaction as defined in our equations. 
 
 There are several interesting features of these  results: (i) the modification of scale factor dynamics is not significant in comparison to the fixed scale factor case $a=t^{1/2}$;  early time behaviour of the scale factor $a(t) \sim t^{0.7}$ is approximately the same for  both the massive and massless scalar field (for the masses shown). This is a reflection of the fact that particle numbers in the modes shown are similar for early times. (ii) There is more particle creation for smaller values of $k$ than larger ones, a feature consistent with the intuition that smaller modes are more readily created. (iii) The massive scalar field case has larger oscillations in particle number for all modes at larger values of the scale factor; this is due to the term $a^3 m_\phi$ which dominates the evolution of $\alpha_k$ in eqn.(\ref{br-tdse}). (iv) Particle production is lowest for the Planck scale modes $k=1$. Again this is an expected feature since larger $k$ particles are harder to produce. 

  \section{Summary and Discussion}   
     
 We defined a self-consistent and non-perturbative Hamiltonian formalism for coupling a quantum scalar field system to a classical FLRW universe. We used this to compute the dynamics of the scale factor by using the evolving mode summed energy density as the quantized source. This demonstrated the viability of the method at least in the setting of homogeneous and isotropic spacetime. 
 
 Our results are derived in the Schrodinger picture for the evolving quantum field. This allowed computation of created particle number at any time during the evolution of the universe rather than just in the asymptotic region. This calculation revealed  the new feature of oscillations in particle number on a fixed background, a result that remains in the self-consistent evolution defined by our coupled classical-quantum equations. Notably, the calculation showed no divergent ``back reaction" on spacetime if the the scale factor is evolved together with the  quantum state from initial data, including the zero mode of the scalar field.  Indeed, as already noted, in our approach there is no apriori background on which a quantum field back reacts, so the term ``backreaction" loses its meaning.  We note also that the evolving state of each mode was restricted to the Gaussian form, but with evolving width. This simplified the problem to a set of coupled ordinary differential equations,  one for each mode.  Removing this restriction would require solving a functional differential equation. 
 
 These results provide a proof of concept for computing self-consistent and non-perturbative evolution of quantum matter and classical spacetime in the first order formalism. It is readily generalized to field theoretic systems without first class constraints. With constraints that generate evolution beyond homogeneity, the consistency of the system in the sense that constraints are preserved if matter contributions to the constraints are replaced by expectation values is an interesting question. Intuitively it seems the answer is that the system remains consistent for the reason that  the gravitational contributions to the constraints, being classical, retain their algebra, while the matter terms contain gravity phase space variables in the same functional forms as in the classical theory.  

The formalism we have described may be applied to field theoretic systems. Among these is the charged scalar-electromagnetic field theory where the charged scalar field is quantized. This would address the ``back reaction" problem in Schwinger effect. Another is the problem of back reaction of Hawking radiation on the geometry of a Schwarzschild black hole: in our approach it is apparent that evolution of quantum matter is unitary by design so the only question is how the geometry and matter evolve self-consistently from quantum-classical initial data. Work in this direction is in progress.

\section*{Acknowledgements}
This work was supported by the Natural Science and Engineering  Research Council of Canada. S.S. is supported in part by the Young Faculty Incentive Fellowship from IIT Delhi.

   \bibliography{BackreactionFRW}
   
     \end{document}